\documentclass[12pt]{article}
\usepackage{amsmath,amssymb,amsthm,amsxtra,overpic,bbm,bm,epsfig,subfigure}
\usepackage{color}
\usepackage{comment}
\usepackage[all,cmtip]{xy}
\usepackage{cite}

\usepackage{slashed,stmaryrd}

\renewcommand{\thefootnote}{\fnsymbol{footnote}}

\begin{document}

\vspace{0.2cm}

\begin{center}
{\large\bf Broken $S^{}_{3{\rm L}} \times S^{}_{3{\rm R}}$ Flavor Symmetry and Leptonic CP Violation}
\end{center}

\vspace{0.2cm}

\begin{center}
{\bf Zong-guo Si}~$^a$ \footnote{E-mail: zgsi@sdu.edu.cn}, \quad {\bf Xing-hua Yang}~$^a$ \footnote{E-mail: yangxh@mail.sdu.edu.cn}, \quad {\bf Shun Zhou}~$^{b,~c,~d}$ \footnote{E-mail: zhoush@ihep.ac.cn}
\\
{$^a$School of Physics, Shandong University, Jinan, Shandong 250100, China}\\

{$^b$Institute of High Energy Physics, Chinese Academy of
Sciences, Beijing 100049, China \\
$^c$School of Physical Sciences, University of Chinese Academy of Sciences, Beijing 100049, China\\
$^d$Center for High Energy Physics, Peking University, Beijing 100871, China}
\end{center}

\vspace{1.5cm}

\begin{abstract}
In the framework of canonical seesaw model, we present a simple but viable scenario to explicitly break an $S^{}_{3{\rm L}} \times S^{}_{3{\rm R}}$ flavor symmetry in the leptonic sector. It turns out that the leptonic flavor mixing matrix is completely determined by the mass ratios of charged leptons (i.e., $m^{}_e/m^{}_\mu$ and $m^{}_\mu/m^{}_\tau$) and those of light neutrinos (i.e., $m^{}_1/m^{}_2$ and $m^{}_2/m^{}_3$). The latest global-fit results of three neutrino mixing angles $\{\theta^{}_{12}, \theta^{}_{13}, \theta^{}_{23}\}$ and two neutrino mass-squared differences $\{\Delta m^2_{21}, \Delta m^2_{31}\}$ at the $3\sigma$ level are used to constrain the parameter space of $\{m^{}_1/m^{}_2, m^{}_2/m^{}_3\}$. The predictions for the mass spectrum and flavor mixing are highlighted: (1) The neutrino mass spectrum shows a hierarchical pattern and a normal ordering, e.g., $m^{}_1 \approx 2.2~{\rm meV}$, $m^{}_2 \approx 8.8~{\rm meV}$ and $m^{}_3 \approx 52.7~{\rm meV}$; (2) Only the first octant of $\theta^{}_{23}$ is allowed, namely, $41.8^\circ \lesssim \theta^{}_{23} \lesssim 43.3^\circ$; (3) The Dirac CP-violating phase $\delta \approx -22^\circ$ deviates significantly from the maximal value $-90^\circ$. All these predictions are ready to be tested in the ongoing and forthcoming neutrino oscillation experiments. Moreover, we demonstrate that the cosmological matter-antimatter asymmetry can be explained via resonant leptogenesis, including the individual lepton-flavor effects. In our scenario, the leptonic CP violation at low- and high-energy scales are closely connected.
\end{abstract}

\begin{flushleft}
\hspace{0.88cm} PACS number(s):  11.30.Fs, 11.30.Hv, 14.60.Pq
\end{flushleft}

\def\thefootnote{\arabic{footnote}}
\setcounter{footnote}{0}

\newpage

\section{Introduction}

Recent neutrino oscillation experiments have firmly established that neutrinos are massive particles and lepton flavors are significantly mixed~\cite{Kajita:2016cak,McDonald:2016ixn}. The leptonic flavor mixing is described by a $3\times 3$ unitary matrix $V$, namely, the so-called Pontecorvo-Maki-Nakagawa-Sakata (PMNS) matrix~\cite{Pontecorvo:1957cp,Maki:1962mu}, which is conventionally parametrized through three mixing angles $\{\theta^{}_{12}, \theta^{}_{13}, \theta^{}_{23}\}$, one Dirac CP-violating phase $\delta$, and two Majorana CP-violating phases $\{\rho, \sigma\}$. As usual, we take the standard parametrization of the PMNS matrix~\cite{Olive:2016xmw}
\begin{eqnarray}
V = \left( \begin{matrix} c^{}_{13} c^{}_{12} & c^{}_{13} s^{}_{12} & s^{}_{13} e^{-{\rm i}\delta} \cr -s_{12}^{} c_{23}^{} - c_{12}^{} s_{13}^{} s_{23}^{} e^{{\rm i}\delta}_{} & + c_{12}^{} c_{23}^{} - s_{12}^{} s_{13}^{} s_{23}^{} e^{{\rm i}\delta}_{} & c_{13}^{} s_{23}^{} \cr + s_{12}^{} s_{23}^{} - c_{12}^{} s_{13}^{} c_{23}^{} e^{{\rm i}\delta}_{} & - c_{12}^{} s_{23}^{} - s_{12}^{} s_{13}^{} c_{23}^{} e^{{\rm i}\delta}_{} & c_{13}^{} c_{23}^{} \end{matrix} \right) \cdot P^{}_\nu
\end{eqnarray}
where $c^{}_{ij} \equiv \cos \theta^{}_{ij}$ and $s^{}_{ij} \equiv \sin \theta^{}_{ij}$ have been defined, and $P^{}_\nu \equiv {\rm Diag}\{e^{{\rm i}\rho}, e^{{\rm i}\sigma}, 1\}$ is a diagonal matrix of two Majorana CP-violating phases. The latest global analysis of neutrino oscillation data~\cite{Esteban:2016qun} yields the best-fit values of three mixing angles $\theta^{}_{12} \approx 33.6^\circ$, $\theta^{}_{13} \approx 8.5^\circ$ and $\theta^{}_{23} \approx 41.6^\circ$, and those of two neutrino mass-squared differences $\Delta m^2_{21} \equiv m^2_2 - m^2_1 \approx 7.50\times 10^{-5}~{\rm eV}^2$ and $\Delta m^2_{31} \equiv m^2_3 - m^2_1 \approx 2.52 \times 10^{-3}~{\rm eV}^2$. Although the normal ordering of neutrino masses (i.e., $m^{}_1 < m^{}_2 < m^{}_3$) is slightly favored and there is a preliminary hint on a nearly-maximal CP-violating phase $\delta \approx 261^\circ$ or equivalently $\delta \approx -99^\circ$, the possibility of an inverted mass ordering (i.e., $m^{}_3 < m^{}_1 < m^{}_2$) and CP conservation in the leptonic sector have not yet been excluded~\cite{Esteban:2016qun}. See also the independent global-fit results from two other groups~\cite{Capozzi:2016rtj,Capozzi:2017ipn,Forero:2014bxa}. An unambiguous determination of neutrino mass ordering and a robust discovery of leptonic CP violation in neutrino oscillations are two primary goals of future medium-baseline reactor (e.g., JUNO~\cite{An:2015jdp} and RENO-50~\cite{B.Kim:2017yps}) and long-baseline accelerator neutrino experiments (e.g., T2K~\cite{Abe:2011ks}, NO$\nu$A~\cite{Ayres:2004js} and LBNF-DUNE~\cite{Acciarri:2015uup}).

In order to accommodate neutrino masses, we can introduce one right-handed neutrino $N^{}_{\alpha \rm R}$ (for $\alpha = e, \mu, \tau$) for each lepton family and write down the gauge-invariant Lagrangian relevant for lepton masses and flavor mixing as follows
\begin{eqnarray}
-{\cal L}^{}_\ell = \overline{\ell^{}_{\rm L}} Y^{}_l H E^{}_{\rm R} + \overline{\ell^{}_{\rm L}} Y^{}_\nu \tilde{H} N^{}_{\rm R} + \frac{1}{2} \overline{N^{\rm C}_{\rm R}} M^{}_{\rm R} N^{}_{\rm R} + {\rm h.c.} \; ,
\end{eqnarray}
where $\ell^{}_{\rm L}$ and $\tilde{H} \equiv {\rm i}\sigma^{}_2 H^*$ denote respectively the left-handed lepton doublet and the Higgs doublet, $E^{}_{\rm R}$ and $N^{}_{\rm R}$ the right-handed charged-lepton and neutrino singlets. While $Y^{}_l$ and $Y^{}_\nu$ are the $3\times 3$ Yukawa coupling matrices for charged leptons and neutrinos, $M^{}_{\rm R}$ is a symmetric mass matrix for right-handed Majorana neutrinos. After the Higgs field acquires its vacuum expectation value $\langle H \rangle = v \approx 174~{\rm GeV}$, the gauge symmetry is spontaneously broken down, and the charged-lepton mass matrix $M^{}_l \equiv Y^{}_l v$ and Dirac neutrino mass matrix $M^{}_{\rm D} \equiv Y^{}_\nu v$ can be obtained. Since the Majorana mass term of right-handed neutrinos is not subject to the gauge symmetry breaking, its absolute scale ${\cal O}(M^{}_{\rm R})$ could be much higher than the electroweak scale $\Lambda^{}_{\rm EW} \sim 100~{\rm GeV}$, e.g., ${\cal O}(M^{}_{\rm R}) \sim 10^{14}~{\rm GeV}$ close to the scale of grand unified theories. At low energies, one can integrate out heavy right-handed neutrinos, and the effective neutrino mass matrix is then given by the seesaw formula $M^{}_\nu \approx M^{}_{\rm D} M^{-1}_{\rm R} M^{\rm T}_{\rm D}$~\cite{Minkowski:1977sc,Yanagida:1979ss,Gell-Mann:1979ss,Glashow:1979ss,Mohapatra:1979ia}. In this canonical seesaw model, the smallness of left-handed neutrinos can be attributed to the heaviness of right-handed neutrinos, and the lepton flavor mixing arises from the mismatch between the diagonalizations of $M^{}_l$ and $M^{}_\nu$. However, the seesaw mechanism itself has told us nothing about the flavor structures of lepton mass matrices $M^{}_l$, $M^{}_{\rm D}$ and $M^{}_{\rm R}$. Hence the lepton mass spectra and flavor mixing remain to be understood in the canonical seesaw model~\cite{Xing:2011zza}.

One typical approach to constrain the flavor structures is to impose a continuous or discrete flavor symmetry on the generic Lagrangian in the first place, and then a spontaneous or explicit symmetry breaking is introduced to help accommodate realistic fermion mass spectra, flavor mixing angles and CP violation~\cite{Harari:1978yi,Froggatt:1978nt}. In Ref.~\cite{Harari:1978yi}, Harari, Haut and Weyers proposed a specific model for quark mass spectra and flavor mixing based on an $S^{}_{3{\rm L}} \times S^{}_{3{\rm R}}$ symmetry, predicting a flavor democracy in the quark sector. Since then, there have been a great number of theoretical works on how to break the flavor democracy or the $S^{}_{3{\rm L}} \times S^{}_{3{\rm R}}$ symmetry in order to explain mass spectra and flavor mixing in both quark and lepton sectors~\cite{Koide:1989ds, Tanimoto:1989qh, Kaus:1990ij, Branco:1990fj, Fritzsch:1989qm, Fritzsch:1994yx, Branco:1995pw, Fritzsch:1995dj, Xing:1996hi, Fukugita:1998vn, Mondragon:1998gy, Fritzsch:1999ee, Haba:2000rf, Branco:2001hn, Fujii:2002jw, Harrison:2003aw, Kubo:2003iw, Chen:2004rr, Fritzsch:2004xc, Rodejohann:2004qh, Araki:2005ec, Teshima:2005bk, Kimura:2005sx, Mohapatra:2006pu, Koide:2006vs, Mondragon:2007af, Xing:2010iu, Teshima:2011wg, Zhou:2011nu, Dong:2011vb, Dev:2012ns, Canales:2012dr, Jora:2012nw, Canales:2013cga, Arbelaez:2016mhg, Yang:2016esx, Xu:2016arj, Pramanick:2016mdp, Guevara:2017ebg, Fritzsch:2017tyf}. In this paper, we put forward a simple but viable scenario for lepton mass spectra and flavor mixing, in which the $S^{}_{3{\rm L}} \times S^{}_{3{\rm R}}$ symmetry in the lepton sector is explicitly broken. To be explicit, we assume the following lepton mass matrices
\begin{eqnarray}
M^{}_l = M^0_l + \Delta M^{}_l \; , \quad M^{}_{\rm D} = M^0_{\rm D} + \Delta M^{}_{\rm D} \; , \quad M^{}_{\rm R} = M^0_{\rm R} + \Delta M^{}_{\rm R} \; ,
\end{eqnarray}
where $M^0_{\rm x}$ (for ${\rm x} = l, {\rm D}, {\rm R}$) are determined by the $S^{}_{3{\rm L}} \times S^{}_{3{\rm R}}$ flavor symmetry, and the perturbations $\Delta M^{}_{\rm x}$ (for ${\rm x} = l, {\rm D}, {\rm R}$) explicitly break the flavor symmetry and will be specified later. Here the subscripts ``${\rm L}$" and ``${\rm R}$" of the flavor symmetry $S^{}_{3{\rm L}} \times S^{}_{3{\rm R}}$ indicate that the corresponding $S^{}_3$ symmetry transformation is nontrivially acting only on left-handed and right-handed fermion fields, respectively. As shown in Appendix A, it is straightforward to prove that the lepton mass matrices in the symmetry limit are uniquely given by
\begin{eqnarray}
M^0_l = \frac{c^{}_l}{3} \left(\begin{matrix} 1 & 1 & 1 \cr 1 & 1 & 1 \cr 1 & 1 & 1 \end{matrix}\right)\; , \quad M^0_{\rm D} = \frac{c^{}_{\rm D}}{3} \left(\begin{matrix} 1 & 1 & 1 \cr 1 & 1 & 1 \cr 1 & 1 & 1 \end{matrix}\right)\; , \quad M^0_{\rm R} = \frac{c^{}_{\rm R}}{3} \left[ \left(\begin{matrix} 1 & 1 & 1 \cr 1 & 1 & 1 \cr 1 & 1 & 1 \end{matrix}\right) + r^{}_{\rm R} \left(\begin{matrix} 1 & 0 & 0 \cr 0 & 1 & 0 \cr 0 & 0 & 1 \end{matrix}\right)\right]\; ,
\end{eqnarray}
where $c^{}_{\rm x}$ characterizes the absolute mass scale of $M^0_{\rm x}$ (for ${\rm x} = l, {\rm D}, {\rm R}$), and $r^{}_{\rm R}$ is a dimensionless parameter. It is interesting to note that the relative size of $r^{}_{\rm R}$ is able to determine whether the mass spectrum of heavy Majorana neutrinos is hierarchical or degenerate.

The present work differs from the previous ones in two aspects. First, we propose a rather simple form of the perturbation matrices $\Delta M^{}_l$, $\Delta M^{}_{\rm D}$ and $\Delta M^{}_{\rm R}$ such that neutrino masses and flavor mixing can be well accommodated. Interestingly, the PMNS matrix is fully fixed by the ratios of charged-lepton masses and those of neutrino masses, and a hierarchical pattern of neutrino mass spectrum is favored. Moreover, the prediction for the CP-violating phase is $\delta \approx -22^\circ$, which can be confirmed or disproved in future long-baseline accelerator neutrino oscillation experiments. Second, we calculate the CP asymmetries in the decays of heavy Majorana neutrinos in the early Universe, and find that the observed matter-antimatter asymmetry can be successfully explained via thermal leptogenesis~\cite{Fukugita:1986hr}. However, one has to require a mass degeneracy of heavy Majorana neutrinos and implement the mechanism of resonant leptogenesis~\cite{Pilaftsis:1997jf,Pilaftsis:2003gt}. A direct connection between the low-energy CP violation in neutrino oscillations and the high-energy CP asymmetries in heavy Majorana neutrino decays can be established.

The remaining part of our paper is organized as follows. In Section 2, we specify the perturbation terms $\Delta M^{}_{\rm x}$ (for ${\rm x} = l, {\rm D}, {\rm R}$) and explore their implications for lepton mass spectra, flavor mixing and CP violation. Then, the CP asymmetries from the heavy Majorana neutrino decays are calculated in Section 3. After taking account of resonant enhancement in CP asymmetries and individual lepton-flavor effects, we show that the predicted baryon number asymmetry can be compatible with cosmological observations. We finally summarize our main results in Section 4.

\section{Mass Spectra, Flavor Mixing and CP Violation}

\subsection{Analytical Results}
In the symmetry limit, the lepton mass matrices $M^0_l$, $M^0_{\rm D}$ and $M^0_{\rm R}$ are given in Eq.~(4), where the democratic matrix with all matrix elements being one is present. In the Majorana mass matrix of heavy neutrinos, there is an extra term proportional to the identity matrix, which is however not affected by any orthogonal transformation to diagonalize the democratic matrix. As is well known, the democratic matrix can be diagonalized by the following orthogonal matrix
\begin{eqnarray}
V^{}_0 = \frac{1}{\sqrt{6}} \left(\begin{matrix} \sqrt{3} & 1 & \sqrt{2} \cr -\sqrt{3} & 1 & \sqrt{2} \cr 0 & -2 & \sqrt{2} \end{matrix}\right) \; .
\end{eqnarray}
All three matrices $M^0_l$, $M^0_{\rm D}$ and $M^0_{\rm R}$ are to be diagonalized by $V^{}_0$, so is the effective neutrino mass matrix $M^0_\nu = M^0_{\rm D} (M^0_{\rm R})^{-1} (M^0_{\rm D})^{\rm T}$. Thus, the first two generations of charged leptons and light neutrinos are massless, implying no flavor mixing in the limit of exact $S^{}_{3{\rm L}} \times S^{}_{3{\rm R}}$ flavor symmetry.

To account for lepton mass spectra and flavor mixing, we shall introduce the perturbation terms $\Delta M^{}_l$ and $\Delta M^{}_{\rm D}$, which explicitly break the $S^{}_{3{\rm L}} \times S^{}_{3{\rm R}}$ flavor symmetry. Note that one can also break the mass degeneracy of heavy neutrinos by assuming a proper $\Delta M^{}_{\rm R}$, which is not necessary for our discussions on low-energy phenomenology. We find that simple diagonal perturbations will suffice for our purpose:
\begin{eqnarray}
\Delta M^{}_l = \frac{c^{}_l}{3} \left(\begin{matrix} +{\rm i} \delta^{}_l & 0 & 0 \cr 0 & -{\rm i}\delta^{}_l & 0 \cr 0 & 0 & \varepsilon^{}_l \end{matrix}\right) \; , \quad \Delta M^{}_{\rm D} = \frac{c^{}_{\rm D}}{3} \left(\begin{matrix} 0 & 0 & 0 \cr 0 & -\delta^{}_{\rm D} & 0 \cr 0 & 0 & \varepsilon^{}_{\rm D} \end{matrix}\right) \; ,
\end{eqnarray}
where $0 < \delta^{}_l \ll \varepsilon^{}_l \ll 1$ and $0 < \delta^{}_{\rm D} \ll \varepsilon^{}_{\rm D} \ll 1$ are implied. Some discussions on different forms of perturbations can be found in Ref.~\cite{Yang:2016esx}. Then, we separately diagonalize the charged-lepton and neutrino mass matrices.
\begin{itemize}
\item {\it Charged-lepton mass matrix $M^{}_l$}---It is convenient to work in the hierarchical basis, which is defined by the transformation $V^{\rm T}_0 M^{}_l V^{}_0 = M^\prime_l$. In this basis, one has to further diagonalize the symmetric matrix below~\cite{Fritzsch:1995dj, Fritzsch:2004xc, Xing:2010iu}
    \begin{eqnarray}
    M^\prime_l = \frac{c^{}_l}{9} \left( \begin{matrix} 0 & {\rm i}\sqrt{3} \delta^{}_l & {\rm i}\sqrt{6}\delta^{}_l \cr {\rm i}\sqrt{3} \delta^{}_l & 2 \varepsilon^{}_l & -\sqrt{2} \varepsilon^{}_l \cr {\rm i}\sqrt{6} \delta^{}_l & -\sqrt{2} \varepsilon^{}_l & 9 + \varepsilon^{}_l \end{matrix}\right) \; .
    \end{eqnarray}
    It is worth mentioning that the perturbation matrix $\Delta M^{}_l$ leads not only to nonzero electron and muon masses, but also to leptonic CP violation at both low- and high-energy scales. Based on the relation of a strong hierarchy $\delta^{}_l \ll \varepsilon^{}_l \ll 1$, we can diagonalize $M^\prime_l$ and obtain the masses of charged leptons
    \begin{eqnarray}
    m^{}_e \approx \frac{\delta^2_l}{6\varepsilon^{}_l} c^{}_l \; , \quad m^{}_\mu \approx \frac{2\varepsilon^{}_l}{9} c^{}_l \; , \quad m^{}_\tau \approx c^{}_l \; .
    \end{eqnarray}
    In addition, the unitary matrix $V^{}_l$, diagonalizing the total charged-lepton mass matrix via $V^\dagger_l M^{}_l V^*_l = {\rm Diag}\{m^{}_e, m^{}_\mu, m^{}_\tau\}$, is found to be
    \begin{eqnarray}
    V^{}_l \approx V^{}_0 + \frac{{\rm i}}{\sqrt{6}} \sqrt{\frac{m^{}_e}{m^{}_\mu}} \left(\begin{matrix} 1 & +\sqrt{3} & 0 \cr 1 & - \sqrt{3} & 0 \cr -2 & 0 & 0 \end{matrix}\right) + \frac{1}{2\sqrt{3}} \frac{m^{}_\mu}{m^{}_\tau} \left( \begin{matrix} 0 & \sqrt{2} & -1 \cr 0 & \sqrt{2} & -1 \cr 0 & \sqrt{2} & 2 \end{matrix}\right) \; .
    \end{eqnarray}
    A salient feature is that all three model parameters $(c^{}_l, \varepsilon^{}_l, \delta^{}_l)$ in the charged-lepton sector are determined by three charged-lepton masses, as shown in Eq.~(8), and the unitary matrix $V^{}_l$ depends only on two mass ratios. By making use of the running charged-lepton masses at the energy scale of $M^{}_Z = 91.2~{\rm GeV}$~\cite{Xing:2007fb, Xing:2011aa}: $m^{}_e \approx 0.486570~{\rm MeV}$, $m^{}_\mu \approx 102.718~{\rm MeV}$ and $m^{}_\tau \approx 1746.17~{\rm MeV}$, we get $m^{}_e/m^{}_\mu \approx 0.00474$ and $m^{}_\mu/m^{}_\tau \approx 0.0588$. From Eq.~(8), one can immediately observe the relations $\varepsilon^{}_l \approx (9/2)\cdot (m^{}_\mu/m^{}_\tau)$ and $\delta^{}_l/\varepsilon^{}_l \approx (2/\sqrt{3})\cdot (m^{}_e/m^{}_\mu)^{1/2}$, and then arrive at $\varepsilon^{}_l \approx 0.265$ and $\delta^{}_l \approx 0.0211$, which are consistent with the assumption that $\varepsilon^{}_l$ and $\delta^{}_l$ can be taken as small perturbation parameters.

\item {\it Effective neutrino mass matrix $M^{}_\nu$}---According to the seesaw formula, the effective neutrino mass matrix is given by
    \begin{eqnarray}
    M^{}_\nu \approx \frac{c^2_{\rm D}}{3 c^{}_{\rm R} r^{}_{\rm R}} \left[ r^{}_{\rm R} \left(\begin{matrix} 1 & 1 & 1 \cr 1 & 1 & 1 \cr 1 & 1 & 1 \end{matrix}\right) + \frac{1}{3} \left( \begin{matrix} 0 & 0 & 0 \cr 0 & 2\delta^2_{\rm D} & \delta^{}_{\rm D} \varepsilon^{}_{\rm D} \cr 0 & \delta^{}_{\rm D} \varepsilon^{}_{\rm D} & 2\varepsilon^2_{\rm D} \end{matrix}\right)\right] \; ,
    \end{eqnarray}
    where $0 < r^{}_{\rm R} \ll \delta^2_{\rm D} \ll \varepsilon^2_{\rm D} \ll 1$ has been assumed. Diagonalizing the effective neutrino mass matrix via $V^\dagger_\nu M^{}_\nu V^*_\nu = {\rm Diag}\{m^{}_1, m^{}_2, m^{}_3\}$ and defining $c^{}_\nu \equiv c^2_{\rm D}/(3 c^{}_{\rm R} r^{}_{\rm R})$, one can find that neutrino masses take on a normal ordering and three mass eigenvalues are
    \begin{eqnarray}
    m^{}_1 \approx r^{}_{\rm R} c^{}_\nu \; , \quad m^{}_2 \approx \frac{1}{2} \delta^2_{\rm D} c^{}_\nu \; , \quad m^{}_3 \approx \frac{2}{3} \varepsilon^2_{\rm D} c^{}_\nu \; ,
    \end{eqnarray}
    and the corresponding unitary matrix is
    \begin{eqnarray}
    V^{}_\nu \approx \left( \begin{matrix} 1 & \displaystyle \frac{m^{}_1}{m^{}_2} & \displaystyle \frac{m^{}_1}{m^{}_3} \cr - \displaystyle \frac{m^{}_1}{m^{}_2} & 1 & \displaystyle \frac{1}{\sqrt{3}} \cdot \sqrt{\frac{m^{}_2}{m^{}_3}} \cr -\displaystyle \frac{m^{}_1}{m^{}_3} & -\displaystyle \frac{1}{\sqrt{3}} \cdot \sqrt{\frac{m^{}_2}{m^{}_3}}  & 1 \end{matrix}\right) \; ,
    \end{eqnarray}
    where only the leading term of each matrix element is kept. Similar to $V^{}_l$ in the charged-lepton sector, $V^{}_\nu$ is completely fixed by the ratios of mass eigenvalues. Even with precise values of three neutrino masses, one cannot figure out all the model parameters $\{c^{}_{\rm D}, \delta^{}_{\rm D}, \varepsilon^{}_{\rm D}\}$ and $\{c^{}_{\rm R}, r^{}_{\rm R}\}$ appearing in Eq.~(11). The determination of all these parameters calls for more observables related to heavy Majorana neutrinos, such as the baryon number asymmetry to be discussed in the next section.
\end{itemize}

Therefore, the PMNS matrix can be calculated via $V = V^\dagger_l V^{}_\nu$, where the unitary matrices $V^{}_l$ and $V^{}_\nu$ can be found in Eq.~(9) and Eq.~(12), respectively. More explicitly, we have
\begin{eqnarray}
V &\approx& \frac{1}{\sqrt{6}} \left( \begin{matrix} \sqrt{3} & - \sqrt{3} & 0 \cr 1 & 1 & -2 \cr \sqrt{2} & \sqrt{2} & \sqrt{2} \end{matrix}\right) - \frac{\rm i}{\sqrt{6}} \sqrt{\frac{m^{}_e}{m^{}_\mu}} \left( \begin{matrix} 1 & 1 & -2 \cr \sqrt{3} & -\sqrt{3} & 0 \cr 0 & 0 & 0\end{matrix}\right) + \frac{1}{2\sqrt{3}} \frac{m^{}_\mu}{m^{}_\tau} \left( \begin{matrix} 0 & 0 & 0 \cr \sqrt{2} & \sqrt{2} & \sqrt{2} \cr -1 & -1 & 2 \end{matrix}\right) \nonumber \\
&~& + \frac{1}{3\sqrt{2}} \sqrt{\frac{m^{}_2}{m^{}_3}} \left( \begin{matrix} 0 & 0 & -\sqrt{3} \cr 0 & 2 & 1 \cr 0 & -\sqrt{2} & \sqrt{2}\end{matrix} \right) + \frac{1}{\sqrt{6}} \frac{m^{}_1}{m^{}_2} \left( \begin{matrix} \sqrt{3} & \sqrt{3} & 0 \cr -1 & 1 & 0 \cr -\sqrt{2} & \sqrt{2} & 0 \end{matrix}\right) + \frac{1}{\sqrt{6}} \frac{m^{}_1}{m^{}_3} \left( \begin{matrix} 0 & 0 & \sqrt{3} \cr 2 & 0 & 1 \cr -\sqrt{2} & 0 & \sqrt{2} \end{matrix}\right)  \; .
\end{eqnarray}
It is worthwhile to mention that the contribution proportional to $m^{}_1/m^{}_3$ from the neutrino sector has been retained, while the one to $m^{}_e/m^{}_\tau$ from the charged-lepton sector neglected. The main reason is that although neutrino mass spectrum in Eq.~(11) is also hierarchical, the mass hierarchy of neutrinos cannot be as strong as that of charged leptons. This observation will become clearer soon when the PMNS matrix in Eq.~(13) is confronted with the latest neutrino oscillation data. Some comments on the phenomenological implications for three flavor mixing angles, CP-violating phases and neutrino masses are in order:
\begin{itemize}
\item Comparing the PMNS matrix $V$ in Eq.~(13) and the standard parametrization in Eq.~(1), one can observe
    \begin{eqnarray}
    \sin^2 \theta^{}_{13}  = |V^{}_{e3}|^2 \approx \frac{1}{6} \cdot \frac{m^{}_2}{m^{}_3} - \frac{1}{\sqrt{3}} \cdot \frac{m^{}_1}{m^{}_2} \left(\frac{m^{}_2}{m^{}_3}\right)^{3/2} + \frac{1}{2} \left(\frac{m^{}_1}{m^{}_2}\right)^2 \left(\frac{m^{}_2}{m^{}_3}\right)^2+ \frac{2}{3} \cdot \frac{m^{}_e}{m^{}_\mu} \; .
    \end{eqnarray}
    The smallest mixing angle $\theta^{}_{13}$ is mainly determined by the mass ratio $m^{}_2/m^{}_3$, but with sub-leading contributions from both $m^{}_e/m^{}_\mu$ and $m^{}_1/m^{}_2$. For a rough estimate, we neglect all the terms associated with $m^{}_e/m^{}_\mu$ and $m^{}_1/m^{}_2$. Given the best-fit value $\theta^{}_{13} \approx 8.5^\circ$ or $\sin^2 \theta^{}_{13} \approx 0.022$ from the latest neutrino oscillation data, we obtain $m^{}_2/m^{}_3 \approx 0.132$. However, as $\theta^{}_{13}$ itself is very small, the neglected terms may have a significant impact on the determination of $m^{}_2/m^{}_3$. For instance, if $m^{}_1/m^{}_2 \approx 0.25$ is assumed, one can get $m^{}_2/m^{}_3 \approx 0.167$ by numerically solving Eq.~(14) with the best-fit value $\sin^2 \theta^{}_{13} \approx 0.022$.
\item Then we proceed to calculate the other two mixing angles $\theta^{}_{12}$ and $\theta^{}_{23}$. Adopting the standard parametrization, we have
    \begin{eqnarray}
    \sin^2 \theta^{}_{12} &=& \frac{|V^{}_{e2}|^2}{1-|V^{}_{e3}|^2}  \approx  \frac{1}{2} \left(1 - \frac{m^{}_1}{m^{}_2} \right)^2 \left(1+\frac{1}{6} \cdot \frac{m^{}_2}{m^{}_3}\right)\; , \\
    \sin^2 \theta^{}_{23} &=& \frac{|V^{}_{\mu3}|^2}{1-|V^{}_{e3}|^2} \approx \frac{2}{3} \left[1 - \frac{1}{2\sqrt{3}} \cdot \sqrt{\frac{m^{}_2}{m^{}_3}} - \frac{1}{2} \left(\frac{m^{}_\mu}{m^{}_\tau} + \frac{m^{}_1}{m^{}_3}\right) \right]^2 \left(1+\frac{1}{6} \cdot \frac{m^{}_2}{m^{}_3}\right)\; ,
    \end{eqnarray}
where only the dominant term of $\sin^2 \theta^{}_{13}$ in Eq.~(14) is considered. As indicated by Eq.~(15), $\sin^2 \theta^{}_{12}$ is more sensitive to the mass ratio $m^{}_1/m^{}_2$ than the other one $m^{}_2/m^{}_3$. However, from Eq.~(16), one can observe that the opposite is true for $\sin^2 \theta^{}_{23}$. To estimate the predictions for $\theta^{}_{12}$ and $\theta^{}_{23}$, we input $m^{}_1/m^{}_2 = 0.25$ and $m^{}_2/m^{}_3 = 0.167$, and then find $\theta^{}_{12} \approx 32.5^\circ$ or $\sin^2 \theta^{}_{12} \approx 0.289$ from Eq.~(15) and $\theta^{}_{23} \approx 43.5^\circ$ or $\sin^2 \theta^{}_{23} \approx 0.474$ from Eq.~(16). Both values of $\theta^{}_{12}$ and $\theta^{}_{23}$ are well lying in the allowed ranges $31.38^\circ \leq \theta^{}_{12} \leq 35.99^\circ$ and $38.4^\circ \leq \theta^{}_{23} \leq 52.8^\circ$ at the $3\sigma$ level.

\item Now let us look at the CP-violating phases from the PMNS matrix. Since the complex term in the PMNS matrix is proportional to $(m^{}_e/m^{}_\mu)^{1/2} \approx 0.069$, as shown in Eq.~(13), it is important for $V^{}_{e3}$ but negligible for other matrix elements. Therefore, the Dirac-type CP-violating phase is approximately given as
    \begin{eqnarray}
    \delta \approx -\arctan\left[2 \sqrt{\frac{m^{}_e}{m^{}_\mu}} \cdot \sqrt{\frac{m^{}_3}{m^{}_2}} \left(1 - \sqrt{3} \cdot \frac{m^{}_1}{m^{}_2} \sqrt{\frac{m^{}_2}{m^{}_3}}\right)^{-1}\right] \approx -22.3^\circ \; ,
    \end{eqnarray}
where $m^{}_1/m^{}_2 = 0.25$ and $m^{}_2/m^{}_3 = 0.167$ have been used. Recent measurements from T2K and NO$\nu$A have shown a preliminary hint on $\delta \approx - 90^\circ$. The future long-baseline accelerator neutrino oscillation experiments, such as LBNF-DUNE~\cite{Acciarri:2015uup} and T2HK~\cite{Abe:2014oxa}, and neutrino super-beam experiments, like ESS$\nu$SB~\cite{Wildner:2015yaa} and MOMENT~\cite{Cao:2014bea, Blennow:2015cmn}, are promising to unambiguously discover leptonic CP violation in neutrino oscillations.

Then, considering the freedom of rephasing the charged-lepton fields, we can also extract two Majorana-type CP-violating phases
\begin{eqnarray}
\rho &\approx& - \arctan\left[\frac{1}{\sqrt{3}}\sqrt{\frac{m^{}_e}{m^{}_\mu}} \left(1+\frac{m^{}_1}{m^{}_2}\right)^{-1}\right] \approx -1.8^\circ \; , \nonumber \\
\sigma &\approx& +\arctan\left[\frac{1}{\sqrt{3}}\sqrt{\frac{m^{}_e}{m^{}_\mu}} \left(1-\frac{m^{}_1}{m^{}_2}\right)^{-1}\right] \approx +3.0^\circ\; ,
\end{eqnarray}
where are close to the trivial value $\pi$ or $0$. This is consistent with our previous observation that all the PMNS matrix elements but $V^{}_{e3}$ are real at the leading order.

\item Finally, we come to the neutrino mass spectrum. The best-fit value of the neutrino mass-squared difference $\Delta m^2_{21} \approx 7.50\times 10^{-5}~{\rm eV}^2$, together with $m^{}_1/m^{}_2 \approx 0.25$ and $m^{}_2/m^{}_3 \approx 0.167$, results in a full determination of neutrino masses, i.e.,
    \begin{eqnarray}
    m^{}_1 = \frac{m^{}_1}{m^{}_2} \sqrt{\frac{\Delta m^2_{21}}{1 - (m^{}_1/m^{}_2)^2}} \approx 2.2~{\rm meV} \; , \quad
    m^{}_2 \approx 8.8~{\rm meV} \; , \quad m^{}_3 \approx 52.7~{\rm meV} \; .
    \end{eqnarray}
    On the other hand, the other neutrino mass-squared difference $\Delta m^2_{31} \approx 2.77\times 10^{-3}~{\rm eV}^2$ can be computed by using neutrino masses in Eq.~(19), which is slightly larger than the observed value $2.41\times 10^{-3}~{\rm eV}^2 \lesssim \Delta m^2_{31} \lesssim 2.64\times 10^{-3}~{\rm eV}^2$ in the $3\sigma$ range. In order to see if the ans\"{a}tze of lepton mass matrices in Eqs. (4) and (6) are really consistent with neutrino oscillation data, we shall carry out a complete numerical analysis in the next subsection.
\end{itemize}

Given the perfectly measured charged-lepton masses, the PMNS matrix is now completely fixed by two neutrino mass ratios $m^{}_1/m^{}_2$ and $m^{}_2/m^{}_3$, which can be determined mainly from the observed values of $\sin^2\theta^{}_{12}$ and $\sin^2\theta^{}_{13}$, respectively. Together with the mass-squared difference $\Delta m^2_{21}$, these two neutrino mass ratios can then be used to predict the mixing angle $\theta^{}_{23}$, the CP-violating phase $\delta$, and three neutrino masses $\{m^{}_1, m^{}_2, m^{}_3\}$, which will be soon tested in the future precision data from neutrino oscillation experiments~\cite{Wang:2015rma}. Moreover, the sum of three absolute neutrino masses will be probed with an unprecedented precision by the observations of cosmic microwave background and the large-scale structures~\cite{Abazajian:2016yjj}.

\subsection{Numerical Analysis}
Since a rough estimate of $m^{}_1/m^{}_2$ and $m^{}_2/m^{}_3$ leads to the predictions for neutrino mixing angles and mass-squared differences only in marginal agreement with neutrino oscillation data, a complete numerical calculation is necessary to demonstrate the validity of the ans\"{a}tze of lepton mass matrices in Eqs.~(4) and (6). Now we outline the strategy to carry out our numerical calculations.

First, the values of $m^{}_1/m^{}_2$ and $m^{}_2/m^{}_3$ are randomly chosen from the range $[0, 1]$, as expected for a normal neutrino mass ordering. Then, we extract three mixing angles $\{\theta^{}_{12}, \theta^{}_{13}, \theta^{}_{23}\}$ directly from the PMNS matrix elements, which are functions of $m^{}_1/m^{}_2$ and $m^{}_2/m^{}_3$. On the other hand, the ratio of two neutrino mass-squared differences
\begin{eqnarray}
\frac{\Delta m^2_{21}}{\Delta m^2_{31}} = \frac{m^2_2}{m^2_3} \left(1 - \frac{m^2_1}{m^2_2}\right) \left(1 - \frac{m^2_1}{m^2_2} \cdot \frac{m^2_2}{m^2_3}\right)^{-1} \; ,
\end{eqnarray}
can also be calculated. Both three mixing angles $\{\theta^{}_{12}, \theta^{}_{13}, \theta^{}_{23}\}$ and the ratio $\Delta m^2_{21}/\Delta m^2_{31}$ are then required to be lying within their $3\sigma$ ranges according to the latest global-fit analysis~\cite{Esteban:2016qun}
\begin{eqnarray}
0.271 \leq &\sin^2 \theta^{}_{12}& \leq 0.345 \; , \nonumber \\
0.01934 \leq &\sin^2 \theta^{}_{13}& \leq 0.02392 \; , \\
0.385 \leq &\sin^2 \theta^{}_{23}& \leq 0.635 \; , \nonumber
\end{eqnarray}
for the mixing angles; and
\begin{eqnarray}
7.03\times 10^{-5}~{\rm eV}^2 \leq &\Delta m^2_{21}& \leq 8.09\times 10^{-5}~{\rm eV}^2 \; , \nonumber \\
2.407\times 10^{-3}~{\rm eV}^2 \leq &\Delta m^2_{31}& \leq 2.643\times 10^{-3}~{\rm eV}^2 \; .
\end{eqnarray}
for the neutrino mass-squared differences~\cite{Esteban:2016qun}. In addition, the lightest neutrino mass $m^{}_1$ can be figured out by choosing one value of $\Delta m^2_{21}$ in its $3\sigma$ range, and then the other mass-squared difference $\Delta m^2_{31}$ is computed and confronted with Eq.~(22). Finally, the viable values of $m^{}_1/m^{}_2$ and $m^{}_2/m^{}_3$ satisfying the above requirements are recorded, and the corresponding predictions for three mixing angles, the CP-violating phase $\delta$, and neutrino masses can be obtained.
\begin{figure}[t!]
\begin{center}
\subfigure{
\includegraphics[width=0.52\textwidth]{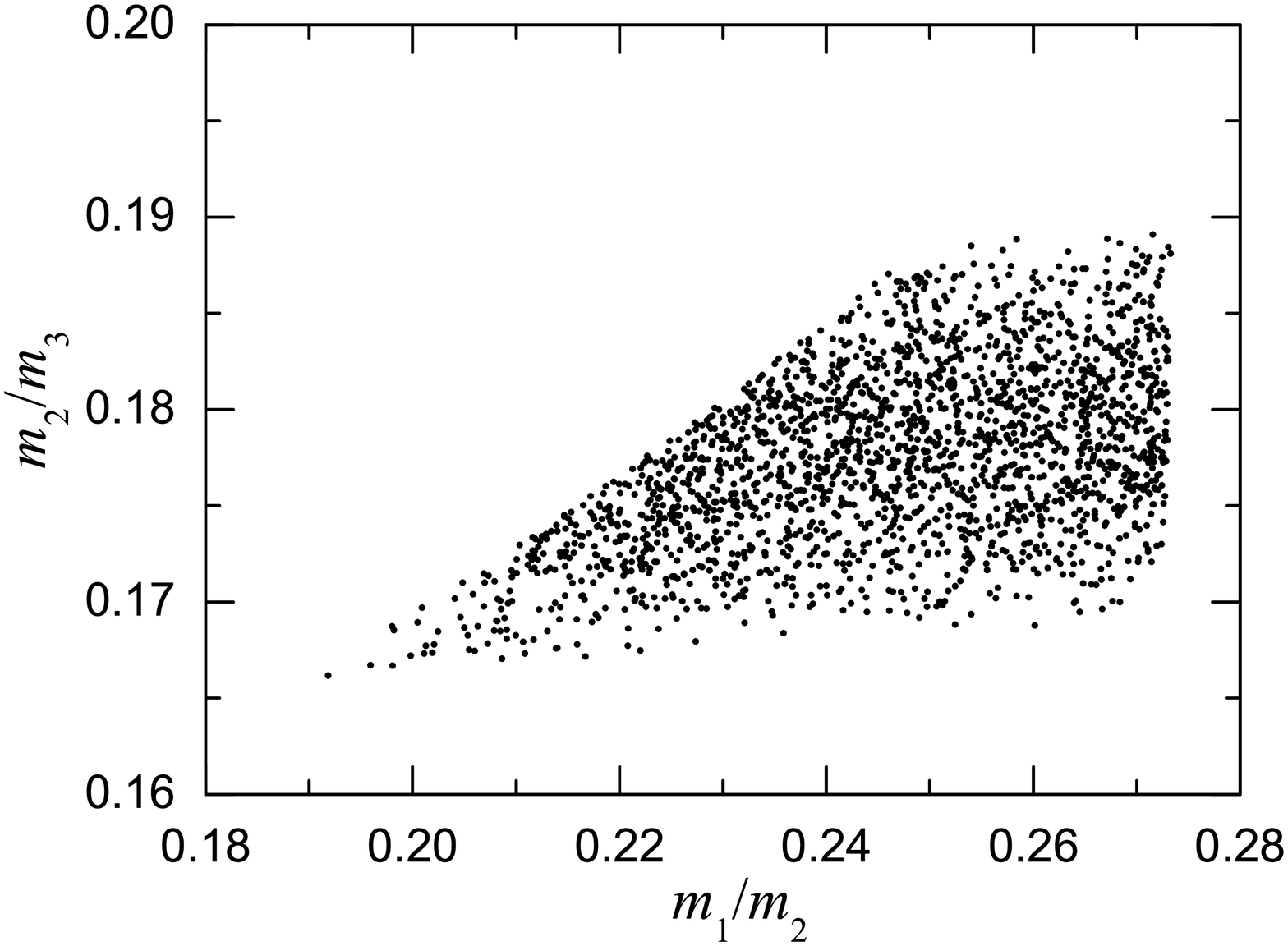} }
\hspace{-1.5cm}
\subfigure{
\includegraphics[width=0.52\textwidth]{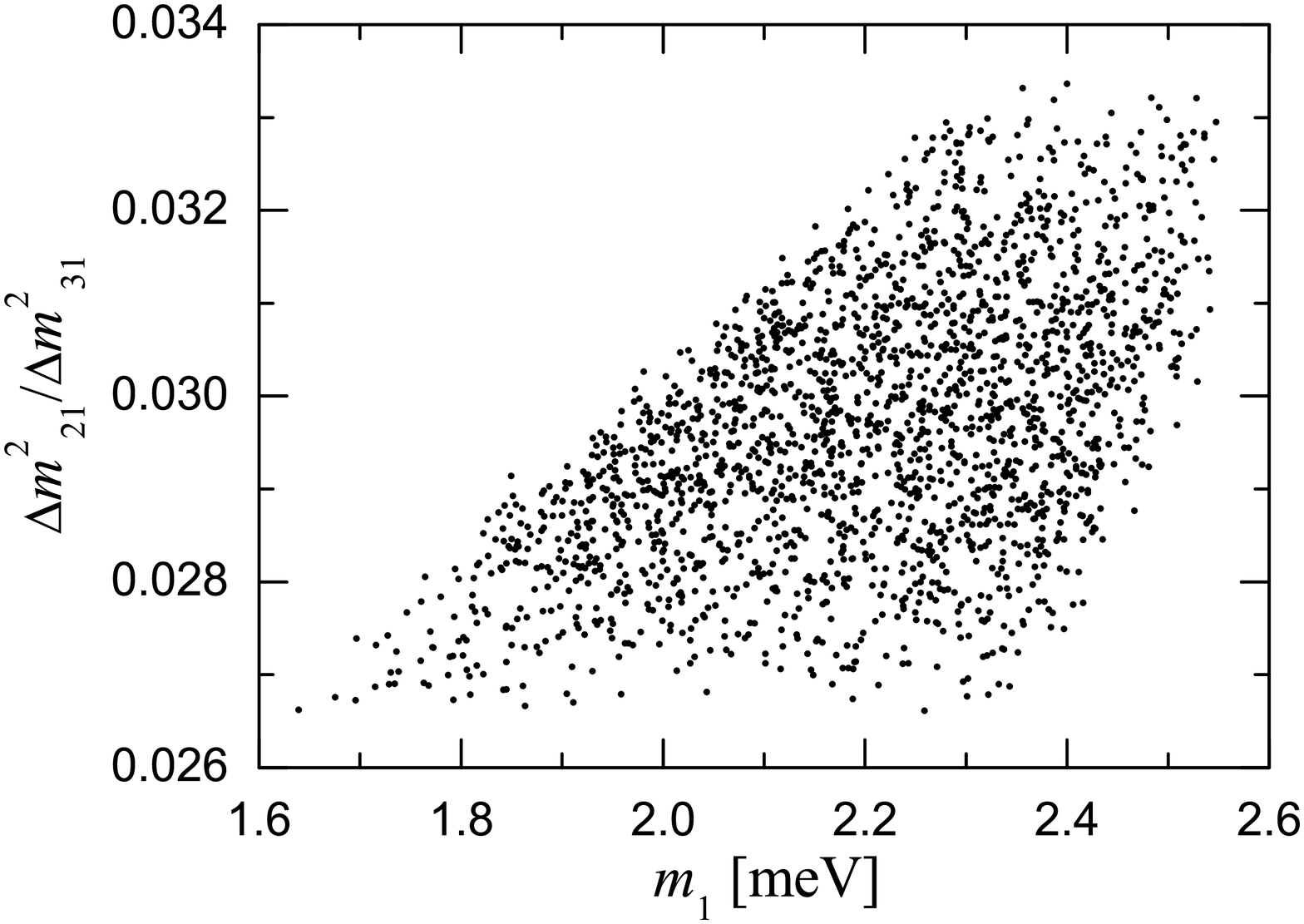} }
\subfigure{
\includegraphics[width=0.52\textwidth]{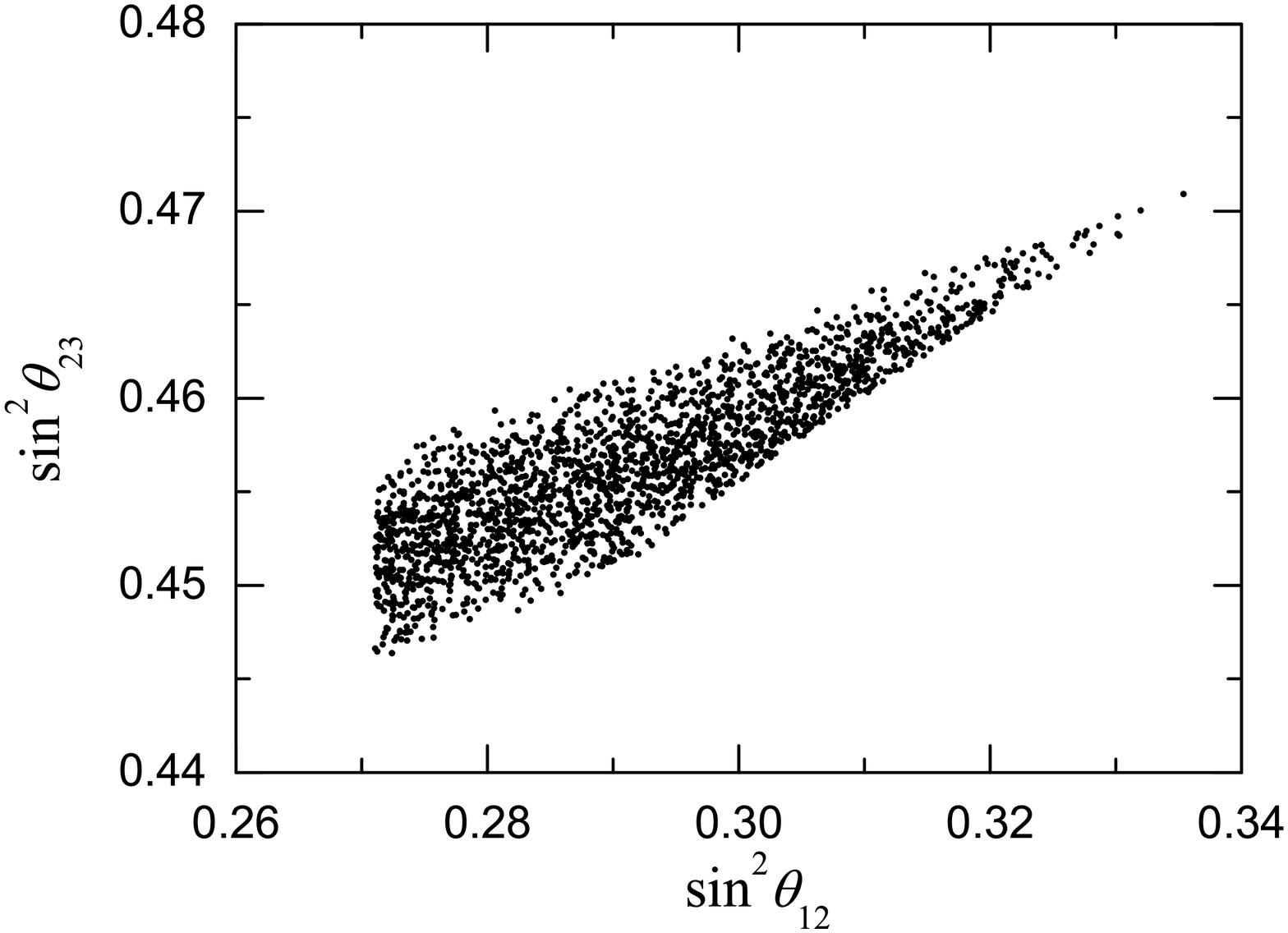} }
\hspace{-1.5cm}
\subfigure{
\includegraphics[width=0.52\textwidth]{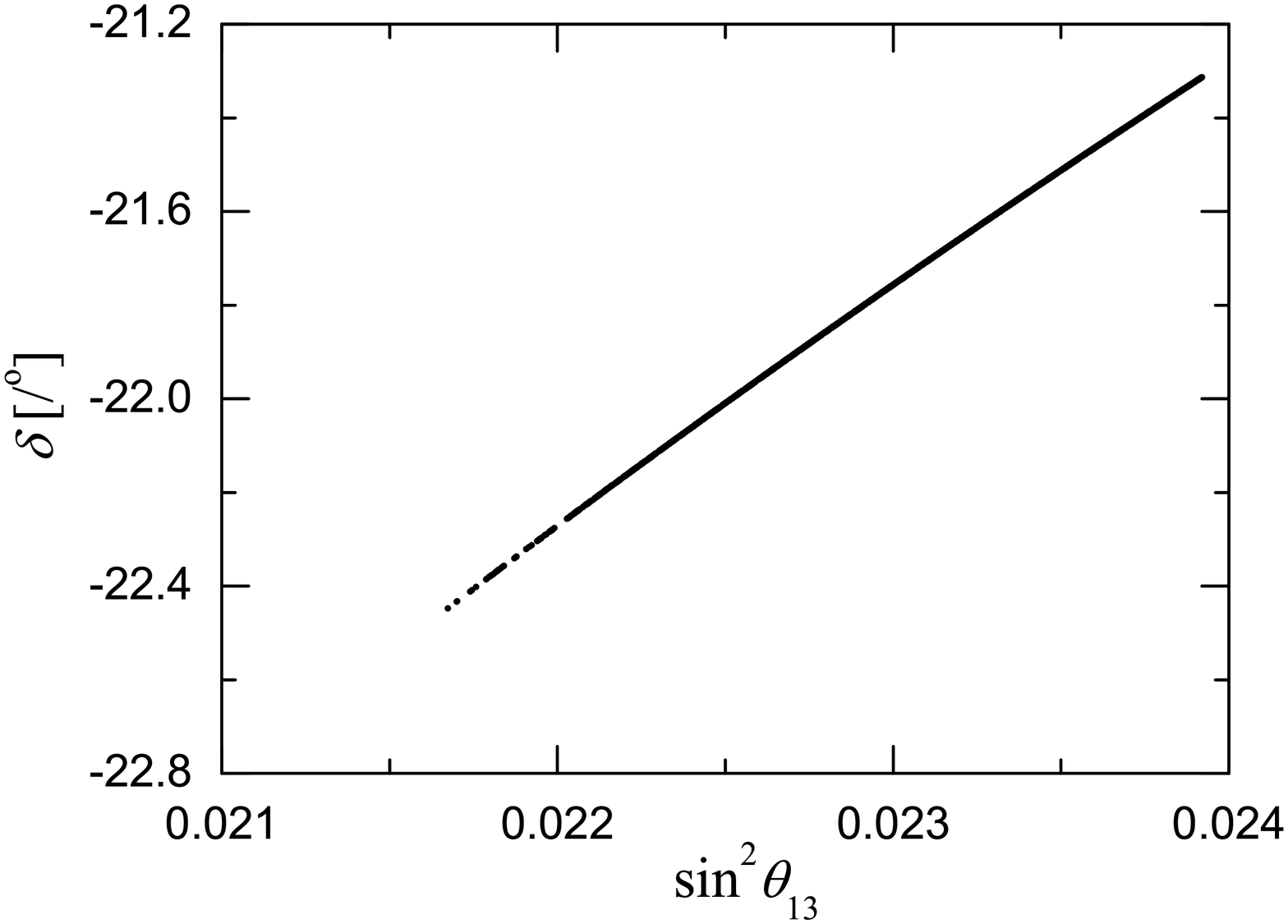} }
\caption{The allowed regions of neutrino mass ratios $(m^{}_1/m^{}_2, m^{}_2/m^{}_3)$, the absolute neutrino mass $m^{}_1$ versus the ratio of two neutrino mass-squared differences $\Delta m^2_{21}/\Delta m^2_{31}$, the mixing parameters $(\sin^2 \theta^{}_{12}, \sin^2 \theta^{}_{23})$ and $(\sin^2 \theta^{}_{13}, \delta)$, where the $3\sigma$ ranges of neutrino mixing parameters in Eqs.~(21) and (22) have been used~\cite{Esteban:2016qun}.}
\end{center}
\end{figure}

The numerical results are given in Fig.~1, where we show the allowed regions of two neutrino mass ratios, three mixing angles, the Dirac CP-violating phase and the absolute neutrino mass. Some comments on the main features are in order.
\begin{itemize}
\item As indicated in the upper-left plot of Fig.~1, only the ranges $0.19\lesssim m^{}_1/m^{}_2 \lesssim 0.27$ and $0.165 \lesssim m^{}_2/m^{}_3 \lesssim 0.185$ are allowed by current neutrino oscillation data. Hence, the neutrino mass spectrum is moderately hierarchical, namely, $m^{}_1 : m^{}_2 : m^{}_3 \approx 1 : 5 : 30$. In constrast, the mass hierarchy of charged leptons is extremely strong, i.e., $m^{}_e : m^{}_\mu : m^{}_\tau \approx 1 : 200 : 3400$. With the help of Eq.~(11), we obtain $r^{}_{\rm R}/\delta^2_{\rm D} \approx m^{}_1/(2m^{}_2) \approx 0.1$ and $\delta^2_{\rm D}/\varepsilon^2_{\rm D} \approx 4m^{}_2/(3m^{}_3) \approx 0.22$, which are in reasonable agreement with the hierarchical limit $r^{}_{\rm R} \ll \delta^2_{\rm D} \ll \varepsilon^2_{\rm D}$.

\item In general, the lightest neutrino mass $m^{}_1$ is allowed to be zero. However, given the best-fit value of $\Delta m^2_{21}/\Delta m^2_{31} \approx 0.0298$, we have $m^{}_2/m^{}_3 = (\Delta m^2_{21}/\Delta m^2_{31})^{1/2} \approx 0.17$ in the limit of $m^{}_1 = 0$, implying $\sin^2\theta^{}_{13} \approx (1/6)\cdot (m^{}_2/m^{}_3) \approx 0.028$ via Eq.~(14). This is obviously in contradiction with the $3\sigma$ upper bound on $\sin^2\theta^{}_{13}$ given in Eq.~(21). As a consequence, a nonzero value of $m^{}_1$ or $m^{}_1/m^{}_2$ is indispensable to reconcile a larger value of $\Delta m^2_{21}/\Delta m^2_{31} \approx 0.0298$ and a smaller value of $\sin^2 \theta^{}_{13}$. In the upper-right plot, we can see the lightest neutrino mass is strictly constrained to $1.6~{\rm meV} \lesssim m^{}_1 \lesssim 2.6~{\rm meV}$, and the other two masses can accordingly be obtained by using two mass ratios.

\item The two plots in the second row of Fig.~1 give us the allowed regions of $(\sin^2 \theta^{}_{12}, \sin^2 \theta^{}_{23})$ and $(\sin^2 \theta^{}_{13}, \delta)$. The first octant of $\theta^{}_{23}$, i.e., $\theta^{}_{23} < 45^\circ$, is favored, and the maximal mixing angle $\theta^{}_{23} = 45^\circ$ is not reachable. Due to the strong correlation, the allowed regions of three mixing angles are severely constrained: $41.8^\circ \lesssim \theta^{}_{23} \lesssim 43.3^\circ$, $31.4^\circ \lesssim \theta^{}_{12} \lesssim 35.5^\circ$, and $8.45^\circ \lesssim \theta^{}_{13} \lesssim 8.90^\circ$. Compared to the $3\sigma$ ranges from current neutrino oscillation data, only a relatively small value of $\theta^{}_{12}$ and a relatively large value of $\theta^{}_{13}$ can survive. Note that the CP-violating phase $\delta \approx -22^\circ$ does not vary much, as it is determined by $m^{}_2/m^{}_3$ and $m^{}_1/m^{}_2$, whose values have already been narrowed down to small regions. All these features are ready to be tested in future neutrino oscillation experiments.
\end{itemize}
Although we have not explored any implications for the effective neutrino mass in tritium beta decays and that in neutrinoless double-beta decays, it is straightforward to estimate them by using the PMNS matrix elements in Eq.~(13) and the corresponding absolute neutrino masses. A normal mass ordering and the hierarchical mass spectrum imply an effective neutrino mass of a few meV, which seems very difficult to be detected in the foreseeable future.

\section{Baryon Number Asymmetry in the Universe}

In this section, we examine whether the lepton mass matrices given in Eqs.~(4) and (6) are also able to explain the baryon number asymmetry in our Universe via thermal leptogenesis~\cite{Fukugita:1986hr}. In particular, we focus on the most attractive scenario in which heavy Majorana neutrinos can be thermally produced in the early Universe, even with a vanishing initial abundance. The central idea of leptogenesis is to first generate lepton number asymmetries from the CP-violating and out-of-equilibrium decays of heavy Majorana neutrinos, which are subsequently converted into baryon number asymmetry via efficient nonperturbative sphaleron processes.  See, e.g., Refs.~\cite{Buchmuller:2004nz, Buchmuller:2005eh, Davidson:2008bu, Hambye:2012fh}, for excellent reviews on recent developments in leptogenesis.

\subsection{Exactly Degenerate Masses}

To calculate the CP asymmetries from the decays of heavy Majorana neutrinos, we should first diagonalize their mass matrix $M^{}_{\rm R}$ to derive the mass spectrum. As seen from the previous section, even with the Majorana mass matrix $M^0_{\rm R}$ in the limit of exact $S^{}_{3{\rm L}}\times S^{}_{3{\rm R}}$ flavor symmetry, one can explain lepton mass spectra, flavor mixing and leptonic CP violation at low energies. Hence, in this subsection, we will not introduce an explicit symmetry-breaking term in the right-handed neutrino sector. In this case, it is evident that $V^\dagger_0 M^{}_{\rm R} V^*_0 = {\rm Diag}\{M^{}_1, M^{}_2, M^{}_3\}$, where $V^{}_0$ is the democratic mixing matrix in Eq.~(5) and three mass eigenvalues are given by
\begin{eqnarray}
M^{}_1 = M^{}_2 = c^{}_{\rm R} r^{}_{\rm R}/3 \; , \quad M^{}_3 = c^{}_{\rm R}(1 + r^{}_{\rm R}/3) \; .
\end{eqnarray}
The mass spectrum is exactly degenerate between $M^{}_1$ and $M^{}_2$, and hierarchical for $r^{}_{\rm R} \ll 1$, i.e., $M^{}_3 \gg M^{}_2 = M^{}_1$. We shall comment on how to break this mass degeneracy in the next subsection.

In thermal leptogenesis, we assume that heavy Majorana neutrinos can be thermally produced at a high temperature, and the lepton asymmetries generated from $N^{}_3$ decays will be completely washed out by the lepton-number-violating processes mediated by two relatively light neutrinos $N^{}_1$ and $N^{}_2$. Therefore, we concentrate only on the CP asymmetries from $N^{}_1$ and $N^{}_2$ decays
\begin{eqnarray}
\epsilon^{}_{i\alpha} \equiv \frac{\Gamma(N^{}_i \to \ell^{}_\alpha H) - \Gamma(N^{}_i \to \overline{\ell}^{}_\alpha H^\dagger)}{\sum_\alpha \left[\Gamma(N^{}_i \to \ell^{}_\alpha H) + \Gamma(N^{}_i \to \overline{\ell}^{}_\alpha H^\dagger)\right]} \; ,
\end{eqnarray}
for $\alpha = e, \mu, \tau$ and $i = 1, 2$, which arise from the interference between the tree and one-loop decay amplitudes. Since the Yukawa interactions of charged leptons are governed by their masses and they become in thermal equilibrium at different temperature, the production and washout of lepton number asymmetries in individual lepton flavors should be taken into account~\cite{Barbieri:1999ma}. It has been found that these lepton-flavor effects could significantly modify the final baryon number asymmetry~\cite{Abada:2006fw,Nardi:2006fx}. This is also the reason why we have to compute the CP asymmetries for all distinct lepton flavors in Eq.~(24).

In the flavor basis where the charged-lepton mass matrix $M^{}_l$ and the heavy Majorana neutrino mass matrix $M^{}_{\rm R}$ are diagonal, the Dirac neutrino mass matrix $\tilde{M}^{}_{\rm D} = V^\dagger_l M^{}_{\rm D} V^*_0$ can be written as
\begin{eqnarray}
\tilde{M}^{}_{\rm D} = c^{}_{\rm D} \left[ \left(\begin{matrix} \displaystyle -\frac{1}{6} \delta^{}_{\rm D} & \displaystyle \frac{1}{6\sqrt{3}} \delta^{}_{\rm D} & \displaystyle \frac{1}{3\sqrt{6}} \delta^{}_{\rm D} \cr \displaystyle \frac{1}{6\sqrt{3}} \delta^{}_{\rm D} & \displaystyle \frac{2}{9} \varepsilon^{}_{\rm D} & -\displaystyle \frac{\sqrt{2}}{9} \varepsilon^{}_{\rm D} \cr \displaystyle \frac{1}{3\sqrt{6}} \delta^{}_{\rm D} & -\displaystyle \frac{\sqrt{2}}{9} \varepsilon^{}_{\rm D} & \displaystyle 1 + \frac{1}{9} \varepsilon^{}_{\rm D} \end{matrix}\right) - {\rm i} \sqrt{\frac{m^{}_e}{m^{}_\mu}} \left( \begin{matrix} \displaystyle \frac{1}{6\sqrt{3}} \delta^{}_{\rm D} & \displaystyle \frac{2}{9} \varepsilon^{}_{\rm D} & \displaystyle -\frac{\sqrt{2}}{9} \varepsilon^{}_{\rm D} \cr \displaystyle -\frac{1}{6} \delta^{}_{\rm D} & \displaystyle \frac{1}{6\sqrt{3}} \delta^{}_{\rm D} & \displaystyle \frac{1}{3\sqrt{6}} \delta^{}_{\rm D} \cr 0 & 0 & 0 \end{matrix}\right)\right] \; , ~~~~~
\end{eqnarray}
where $m^{}_\mu/m^{}_\tau \ll \delta^{}_{\rm D} \ll \varepsilon^{}_{\rm D} \ll 1$ is assumed and only the leading-order terms are retained. As the imaginary parts of the matrix elements in $\tilde{M}^{}_{\rm D}$ are suppressed by $(m^{}_e/m^{}_\mu)^{1/2} \approx 0.069$, compared to the real parts, small CP asymmetries are generally expected. Taking account of both self-energy and vertex corrections to heavy Majorana neutrino decays, one can get~\cite{Davidson:2008bu, Xing:2011zza}
\begin{eqnarray}
\epsilon^{}_{i\alpha} = \frac{1}{8\pi v^2 {\cal H}^{}_{ii}} \sum_{k\neq i}  \left\{ {\rm Im} \left[(\tilde{M}^{}_{\rm D})^*_{\alpha i} (\tilde{M}^{}_{\rm D})^{}_{\alpha k} {\cal H}^{}_{ik}\right] F(x^{}_{ki}) + {\rm Im} \left[(\tilde{M}^{}_{\rm D})^*_{\alpha i} (\tilde{M}^{}_{\rm D})^{}_{\alpha k} {\cal H}^*_{ik}\right] G(x^{}_{ki}) \right\}\; ,
\end{eqnarray}
with ${\cal H} \equiv \tilde{M}^\dagger_{\rm D} \tilde{M}^{}_{\rm D}$ and $x^{}_{ki} \equiv M^2_k/M^2_i$. The loop functions are $G(x) \equiv 1/(1-x)$ and
\begin{eqnarray}
F(x) \equiv \sqrt{x} \left[1 + \frac{1}{1-x} + (1+x) \ln\left(\frac{x}{1+x}\right)\right] \; .
\end{eqnarray}
Some discussions on the CP asymmetries in Eq.~(26) are helpful. First, it should be noticed that the total CP asymmetries $\epsilon^{}_i = \sum_\alpha \epsilon^{}_{i\alpha} \propto {\rm Im}\left[{\cal H}^2_{ik}\right]$ are vanishing due to the fact that ${\cal H} \equiv \tilde{M}^\dagger_{\rm D} \tilde{M}^{}_{\rm D} = V^{\rm T}_0 M^\dagger_{\rm D} M^{}_{\rm D} V^{}_0$ is a real matrix. Consequently, CP asymmetry exists only for each lepton flavor, and one has to study the generation and evolution of lepton number asymmetry of each lepton flavor. In addition, as shown in Eq.~(25), the matrix elements of $\tilde{M}^{}_{\rm D}$ in the third row are real, so the tau-flavor asymmetries are zero. Second, the exact mass degeneracy requires a more careful treatment of the would-be singularity in $1/(M^2_2-M^2_1)$ residing in the loop functions. In fact, the contribution to CP asymmetries from the self-energy diagram is vanishing for $M^{}_1 = M^{}_2$. Third, In the limit of $M^{}_3 \gg M^{}_2 = M^{}_1$, the loop function can be simplified as $F(x) \to -3/(2\sqrt{x})$ and $G(x) \to -1/x$ for $x \to +\infty$. With the help of Eqs.~(25) and (26), we can obtain
\begin{eqnarray}
\epsilon^{}_{1e} &=& - \epsilon^{}_{1\mu} \approx -\frac{m^{}_1 M^{}_1}{48\pi v^2} \left[ \sqrt{\frac{m^{}_2}{m^{}_1}} (1-2\ln 2) - \frac{3\sqrt{6}}{2} \cdot \sqrt{\frac{M^{}_1}{M^{}_3}} \right]\sqrt{\frac{m^{}_3}{m^{}_1}} \cdot \sqrt{\frac{m^{}_e}{m^{}_\mu}} \; , \nonumber \\
\epsilon^{}_{2e} &=& - \epsilon^{}_{2\mu} \approx +\frac{m^{}_1 M^{}_1}{48\pi v^2} \left[ \sqrt{\frac{m^{}_2}{m^{}_1}} \cdot \sqrt{\frac{m^{}_2}{m^{}_3}} (1-2\ln 2) + \frac{9}{\sqrt{2}} \cdot \sqrt{\frac{M^{}_1}{M^{}_3}} \right]\sqrt{\frac{m^{}_2}{m^{}_1}} \cdot \sqrt{\frac{m^{}_e}{m^{}_\mu}} \; ,
\end{eqnarray}
and $\epsilon^{}_{1\tau} = \epsilon^{}_{2\tau} = 0$, where the relations in Eq.~(11) and (23) have been implemented to express the parameters $\{c^{}_{\rm D}, \delta^{}_{\rm D}, \varepsilon^{}_{\rm D}\}$ in terms of mass ratios of light and heavy Majorana neutrinos. Unfortunately, it turns out that the CP asymmetries in Eq.~(28) are not sufficient to achieve a successful leptogenesis. The main reasons are summarized below:
\begin{enumerate}
\item As we have mentioned before, the total CP asymmetry in $N^{}_i$ decays is vanishing, namely, $\epsilon^{}_i \equiv \epsilon^{}_{ie} + \epsilon^{}_{i\mu} + \epsilon^{}_{i\tau}  = 0$ for $i = 1, 2$. If $M^{}_1 \gtrsim 10^{12}~{\rm GeV}$, the one-flavor treatment of leptogenesis is valid and only the total CP asymmetry $\epsilon^{}_i$ matters. For $10^9~{\rm GeV} \lesssim M^{}_1 \lesssim 10^{12}~{\rm GeV}$, the tau-flavor Yukawa interaction of charged leptons becomes in thermal equilibrium, so we have to deal with tau flavor and the other lepton flavors separately. However, the CP asymmetries $\epsilon^{}_i = 0$ in the former case, and $\epsilon^{}_{i\tau} = 0$ and $\epsilon^{}_{ie} + \epsilon^{}_{i\mu} = 0$ in the latter case, indicate that no lepton number asymmetries can be produced.

\item For $M^{}_1 \lesssim 10^9~{\rm GeV}$, both tau- and muon-flavor Yukawa interactions of charged leptons are in thermal equilibrium, and thus we have to distinguish the production and washout of lepton asymmetries in electron and muon flavors, which will be converted into baryon number asymmetry via sphaleron processes. Since the source of CP violation comes in with a factor of $(m^{}_e/m^{}_\mu)^{1/2}$, both the Dirac CP-violating phase in Eq.~(17) and CP asymmetries in Eq.~(28) are suppressed. For an order-of-magnitude estimate, we have
    \begin{eqnarray}
    |\epsilon^{}_{ie}| = |\epsilon^{}_{i\mu}| \approx \frac{m^{}_1 M^{}_1}{48\pi v^2} \sqrt{\frac{m^{}_e}{m^{}_\mu}} \lesssim 3 \times 10^{-11} \; ,
    \end{eqnarray}
    where $m^{}_1 \approx 2.2~{\rm meV}$ and $M^{}_1 \lesssim 10^9~{\rm GeV}$ have been used. These CP asymmetries are too small to account for the observed baryon-to-photon ratio $\eta^{}_{\rm B} \equiv n^{}_{\rm B}/n^{}_\gamma = (6.09\pm 0.06)\times 10^{-10}$ at the $95\%$ confidence level~\cite{Ade:2015xua}.
\end{enumerate}
In summary, the flavor structures of lepton mass matrices in Eqs.~(4) and (6) cause zero total CP asymmetries, but this is not the case for the CP asymmetries of individual lepton flavors. For the flavor effects to be at work, the lightest heavy neutrino mass is bounded from above, i.e., $M^{}_1 \lesssim 10^9~{\rm GeV}$, leading to insufficient production of lepton number asymmetries. In the next subsection, we shall go beyond the scenario of the partially degenerate mass spectrum $M^{}_1 = M^{}_2 \ll M^{}_3$, and consider a tiny mass splitting between $M^{}_1$ and $M^{}_2$.

\subsection{Nearly Degenerate Masses}

If the $S^{}_{3{\rm L}} \times S^{}_{3{\rm R}}$ symmetry is also explicitly broken down in the right-handed neutrino sector, the mass degeneracy between $M^{}_1$ and $M^{}_2$ will be shifted. For illustration, we take the following perturbations for the heavy Majorana neutrino mass matrix:
\begin{eqnarray}
\Delta M^{}_{\rm R} = \frac{c^{}_{\rm R}}{3} \left( \begin{matrix} -\delta^{}_{\rm R} & + \delta^{}_{\rm R} & 0 \cr + \delta^{}_{\rm R} & -\delta^{}_{\rm R} & 0 \cr 0 & 0 & 0 \end{matrix}\right) \; ,
\end{eqnarray}
where $|\delta^{}_{\rm R}| \ll 1$ is responsible for the mass splitting between $N^{}_1$ and $N^{}_2$. Even with this nontrivial perturbation matrix, the full mass matrix $M^{}_{\rm R}$ can be diagonalized in the same way as before, i.e., $V^{\rm T}_0 M^{}_{\rm R} V^{}_0 = {\rm Diag}\{M^{}_1, M^{}_2, M^{}_3\}$, where the mass eigenvalues are given by
\begin{eqnarray}
M^{}_1 = c^{}_{\rm R} (r^{}_{\rm R} - 2\delta^{}_{\rm R})/3 \; , \quad M^{}_2 = c^{}_{\rm R} r^{}_{\rm R}/3 \; , \quad M^{}_3 = c^{}_{\rm R}(1 + r^{}_{\rm R}/3) \; .
\end{eqnarray}
The mass degeneracy parameter, defined by $\Delta \equiv (M^{}_2 - M^{}_1)/M^{}_2 = 2\delta^{}_{\rm R}/r^{}_{\rm R}$, can be very small, since it is just the ratio of the symmetry-breaking to symmetry-preserving terms. Another advantage of the perturbations in Eq.~(30) can be immediately recognized: the mass spectrum of three light neutrinos and lepton flavor mixing are not affected at all. Hence, all the previous conclusions on low-energy phenomena are still valid.

As pointed out in Refs.~\cite{Pilaftsis:1997jf, Pilaftsis:2003gt}, the CP asymmetries arising from the self-energy corrections involving $N^{}_1$ and $N^{}_2$ will be greatly enhanced when their masses are nearly degenerate. For this reason, we neglect the minor contributions from the vertex correction, and those from the heaviest neutrino $N^{}_3$ as well, and arrive at~\cite{Xing:2006ms,Zhang:2015tea}
\begin{eqnarray}
\epsilon^{}_{i\alpha} = \frac{1}{8\pi v^2 {\cal H}^{}_{ii}} \sum_{k \neq i} \left\{ {\rm Im} \left[(\tilde{M}^{}_{\rm D})^*_{\alpha i} (\tilde{M}^{}_{\rm D})^{}_{\alpha k} {\cal H}^{}_{ik}\right] \sqrt{x^{}_{ki}} + {\rm Im} \left[(\tilde{M}^{}_{\rm D})^*_{\alpha i} (\tilde{M}^{}_{\rm D})^{}_{\alpha k} {\cal H}^*_{ik}\right] \right\}G^\prime(x^{}_{ki}) \; ,
\end{eqnarray}
where $x^{}_{ki}\equiv M^2_k/M^2_i$ has been defined and the regularized loop function reads
\begin{eqnarray}
G^\prime(x^{}_{ki}) = \frac{1 - x^{}_{ki}}{(1 - x^{}_{ki})^2 + r^2_{ki}} \; .
\end{eqnarray}
Note that the regulator $r^{}_{ki} \equiv \Gamma^{}_k/M^{}_i$, with $\Gamma^{}_k \equiv {\cal H}^{}_{kk}M^{}_k/(8\pi v^2)$ being the total decay width of $N^{}_k$ at the tree level, guarantees the correct behavior of the CP asymmetries in the limit of exact mass degeneracy, i.e., $\epsilon^{}_{i\alpha} \to 0$ for $M^{}_1 \to M^{}_2$. With the help of Eq.~(25), it is straightforward to derive
\begin{eqnarray}
\epsilon^{}_{1e} &=& - \epsilon^{}_{1\mu} = + \frac{1}{3} \cdot \sqrt{\frac{m^{}_2}{m^{}_3}} \cdot \sqrt{\frac{m^{}_e}{m^{}_\mu}} \cdot \frac{\tilde{r}^{}_{21} \Delta}{\Delta^2 + \tilde{r}^2_{21}} \; , \nonumber \\
\epsilon^{}_{2e} &=& - \epsilon^{}_{2\mu} = + \frac{1}{3} \cdot \sqrt{\frac{m^{}_2}{m^{}_3}} \cdot \sqrt{\frac{m^{}_e}{m^{}_\mu}} \cdot \frac{\tilde{r}^{}_{12} \Delta}{\Delta^2 + \tilde{r}^2_{12}} \; ,
\end{eqnarray}
where $\tilde{r}^{}_{21} \equiv r^{}_{21}/2 \approx M^{}_1 m^{}_3/(16\pi v^2)$ and $\tilde{r}^{}_{12} \equiv r^{}_{12}/2 \approx M^{}_1 m^{}_2/(16\pi v^2)$. Similar to the Dirac CP-violating phase $\delta$, the CP asymmetries are suppressed by the small parameter $(m^{}_e/m^{}_\mu)^{1/2}$. However, they can be resonantly enhanced and the maximum of $\tilde{r}^{}_{21} |\Delta|/(\Delta^2 + \tilde{r}^2_{21})$ or $\tilde{r}^{}_{12} |\Delta|/(\Delta^2 + \tilde{r}^2_{12})$ could reach $1/2$, when $|\Delta| \approx \tilde{r}^{}_{21}$ or $\tilde{r}^{}_{12}$ is satisfied. Note also that the sign of $\Delta$ will be fixed by the observed baryon number asymmetry.

In the strong washout regime, where heavy Majorana neutrinos can be thermally produced, the final baryon number asymmetry is independent of the initial conditions. We assume that the initial abundance of heavy Majorana neutrinos is following the thermal distribution at the temperature much higher than their masses, i.e., $T \gtrsim M^{}_1 \approx M^{}_2$. The lepton number asymmetries generated in heavy neutrino decays will not be completely destroyed by the lepton-number-violating inverse decays and scattering processes. In this case, the final baryon number asymmetry can be approximately computed by~\cite{Buchmuller:2004nz,Buchmuller:2005eh}
\begin{eqnarray}
\eta^{}_{\rm B} \approx -0.96\times 10^{-2} \sum^{}_i \sum^{}_\alpha \epsilon^{}_{i\alpha} \kappa^{}_{i\alpha} \; ,
\end{eqnarray}
where the efficiency factors $\kappa^{}_{i\alpha}$ measure how efficiently the lepton number asymmetries will be destroyed. To accurately calculate the efficiency factors, one should solve a full set of Boltzmann equations for the evolution of lepton number asymmetries in different flavors. However, we instead follow an approximate and analytical approach by introducing the decay parameter $K^{}_i \equiv \tilde{m}^{}_i/m^{}_*$, where the effective neutrino mass is defined as $\tilde{m}^{}_i \equiv {\cal H}^{}_{ii}/M^{}_i$ and the equilibrium neutrino mass is $m^{}_* \approx 1.08\times 10^{-3}~{\rm eV}$. If the individual decay parameter $K^{}_{i\alpha} \equiv K^{}_i |(\tilde{M}^{}_{\rm D})^{}_{\alpha i}|^2/{\cal H}^{}_{ii}$ happens to be in the range $5 \lesssim K^{}_{i\alpha} \lesssim 100$, it is a good approximation that $\kappa^{}_{i\alpha} \approx 0.5/K^{1.2}_{i\alpha}$~\cite{Blanchet:2006dq,Blanchet:2006be}. In the limit of degenerate masses, i.e., $M^{}_1 \approx M^{}_2$, one has to replace $K^{}_{i\alpha}$ by the sum $K^{}_{1\alpha} + K^{}_{2\alpha}$ in calculating the individual efficiency factor $\kappa^{}_{i\alpha}$. In our case, it is easy to derive $\tilde{m}^{}_1 \approx m^{}_2$ and $\tilde{m}^{}_2 \approx m^{}_3$, and $K^{}_{1e}/K^{}_1 \approx 1/2$, $K^{}_{1\mu}/K^{}_1 \approx 1/6$, and $K^{}_{2e}/K^{}_2 \approx m^{}_2/(6m^{}_3)$, $K^{}_{2\mu}/K^{}_2 \approx 2/3$. For illustration, we take $m^{}_2 = 8.8~{\rm meV}$ and $m^{}_3 = 52.7~{\rm meV}$ as in Eq.~(26), and then obtain
\begin{eqnarray}
K^{}_1 \approx \frac{m^{}_2}{m^{}_*} \approx 8.15 \; , \quad K^{}_{1e} \approx 4.07 \; , \quad K^{}_{1\mu} \approx 1.36 \; ,
\end{eqnarray}
and
\begin{eqnarray}
K^{}_2 \approx \frac{m^{}_3}{m^{}_*} \approx 48.8 \; , \quad K^{}_{2e} \approx 1.36 \; , \quad K^{}_{2\mu} \approx 32.5 \; ,
\end{eqnarray}
implying $\kappa^{}_{1e} = \kappa^{}_{2e} \approx 0.5/(K^{}_{1e} + K^{}_{2 e})^{1.2} \approx 6.6\times 10^{-2}$ and $\kappa^{}_{1\mu} = \kappa^{}_{2\mu} \approx 0.5/(K^{}_{1\mu} + K^{}_{2 \mu})^{1.2} \approx 7.3\times 10^{-3}$. Put all together, we finally get
\begin{eqnarray}
\eta^{}_{\rm B} \approx -5.4\times 10^{-6} \cdot  \frac{\tilde{r}^{}_{21} \Delta}{\Delta^2 + \tilde{r}^2_{21}} \left(1 + 0.167\frac{\Delta^2 + \tilde{r}^2_{21}}{\Delta^2 + \tilde{r}^2_{12}}\right)
= \left\{ \begin{array}{cc}
6.3\times 10^{-6} \displaystyle \frac{\tilde{r}^{}_{21}}{|\Delta|}, & |\Delta| \gg \tilde{r}^{}_{21} \\
~ & ~ \\
3.8\times 10^{-5} \displaystyle \frac{|\Delta|}{\tilde{r}^{}_{21}}\; , & |\Delta| \ll \tilde{r}^{}_{21}
\end{array}
\right. \; ,
\end{eqnarray}
where $\Delta$ should be negative in order to account for the observed positive $\eta^{}_{\rm B}$, and the parameter
\begin{eqnarray}
\tilde{r}^{}_{21} \approx 3.5\times 10^{-14} \left(\frac{M^{}_1}{10^3~{\rm GeV}}\right) \; ,
\end{eqnarray}
depends crucially on the heavy Majorana neutrino mass $M^{}_1$. For $M^{}_1 = 1~{\rm TeV}$, the best-fit value $\eta^{}_{\rm B} = 6.09\times 10^{-10}$ requires the mass degeneracy parameter to be $\Delta = -3.6\times 10^{-10}$ or $-5.6\times 10^{-19}$. As $\Delta = 2\delta^{}_{\rm R}/r^{}_{\rm R}$ arises from the soft breaking of $S^{}_{3{\rm L}} \times S^{}_{3{\rm R}}$ flavor symmetry, it is naturally expected to be small. If $M^{}_1 \approx 10^9~{\rm GeV}$ is assumed, we shall obtain $\Delta = -3.6\times 10^{-4}$ or $-5.6\times 10^{-13}$. Therefore, we have demonstrated that the baryon number asymmetry can also be explained in our scenario by implementing the mechanism of resonant leptogenesis, including lepton flavor effects. It is worthwhile to stress that the CP-violating phase $\delta \approx -22^\circ$ leads to a wrong sign of baryon number asymmetry, which however can be corrected by the degeneracy parameter $\Delta$.

\section{Summary}

In this paper, we have examined a simple but viable scenario to explicitly break the $S^{}_{3{\rm L}} \times S^{}_{3{\rm R}}$ flavor symmetry in the canonical seesaw model. In the symmetry limit, both charged-lepton mass matrix $M^{}_l$ and Dirac neutrino mass matrix $M^{}_{\rm D}$ take on the democratic form, while the heavy Majorana neutrino mass matrix $M^{}_{\rm R}$ consists of a democratic part and another one proportional to the identity matrix. After introducing diagonal perturbation matrices $\Delta M^{}_l$ and $\Delta M^{}_{\rm D}$, we have explored their implications for lepton mass spectra, flavor mixing angles and CP-violating phases at low energies, and calculate the baryon number asymmetry in our Universe via the mechanism of thermal leptogenesis.

At the low-energy scale, the effective neutrino mass matrix is given by the famous seesaw formula $M^{}_\nu = M^{}_{\rm D} M^{-1}_{\rm R} M^{\rm T}_{\rm D}$. The leptonic flavor mixing matrix $V$, arising from the diagonalizations of $M^{}_l$ and $M^{}_\nu$, is completely determined by the mass ratios of charged leptons, i.e., $m^{}_e/m^{}_\mu$ and $m^{}_\mu/m^{}_\tau$, and those of neutrinos, i.e., $m^{}_1/m^{}_2$ and $m^{}_2/m^{}_3$. The $3\sigma$ ranges of three neutrino mixing angles and two mass-squared differences from the latest global analysis of neutrino oscillation data have been implemented to constrain the parameter space of neutrino mass ratios. It turns out that our scenario is well consistent with current neutrino oscillation data, and neutrino mass ratios are found to be $0.19\lesssim m^{}_1/m^{}_2 \lesssim 0.27$ and $0.165 \lesssim m^{}_2/m^{}_3 \lesssim 0.185$. Consequently, a hierarchical pattern of neutrino mass spectrum (e.g., $m^{}_1 \approx 2.2~{\rm meV}$, $m^{}_2 \approx 8.8~{\rm meV}$ and $m^{}_3 \approx 52.7~{\rm meV}$) and a normal mass ordering are favored. The allowed regions of mixing angles are $41.8^\circ \lesssim \theta^{}_{23} \lesssim 43.3^\circ$, $31.4^\circ \lesssim \theta^{}_{12} \lesssim 35.5^\circ$, and $8.45^\circ \lesssim \theta^{}_{13} \lesssim 8.90^\circ$, together with the Dirac CP-violating phase $\delta \approx -22^\circ$, are ready to be tested in future neutrino oscillation experiments.

If the $S^{}_{3{\rm L}} \times S^{}_{3{\rm R}}$ symmetry is preserved in the right-handed neutrino sector, two heavy neutrinos $N^{}_1$ and $N^{}_2$ are exactly degenerate in masses. As a consequence of the flavor structures of lepton mass matrices, the CP asymmetries in heavy neutrino decays are found to be $\epsilon^{}_{i\tau} = 0$ and $\epsilon^{}_{ie} + \epsilon^{}_{i\mu} = 0$ for $i = 1, 2$. An upper bound $M^{}_1 = M^{}_2 \lesssim 10^9~{\rm GeV}$ should be met for the flavored leptogenesis to work efficiently. However, we find that the CP asymmetries for $M^{}_1 = M^{}_2 \lesssim 10^9~{\rm GeV}$ are insufficient to explain the observed baryon number asymmetry. If the flavor symmetry is explicitly broken as well for heavy Majorana neutrinos, the CP asymmetries from the mixing between $N^{}_1$ and $N^{}_2$, which are nearly degenerate in masses, will be resonantly enhanced. In this case, even for $M^{}_1 \approx M^{}_2 \approx 1~{\rm TeV}$, we can successfully explain the observed baryon-to-photon number ratio $\eta^{}_{\rm B} \approx 6.09\times 10^{-10}$ by setting a tiny mass degeneracy $\Delta \equiv (M^{}_2 - M^{}_1)/M^{}_2 \approx -3.6\times 10^{-10}$ or $-5.6 \times 10^{-19}$. As such a mass splitting comes from the flavor symmetry breaking, it is naturally expected to be small.

Notice that the Dirac CP-violating phase, stemming from the perturbation matrix $\Delta M^{}_l$, is also responsible for the CP asymmetries in heavy Majorana neutrino decays, which are indispensable for explaining cosmological matter-antimatter asymmetry. Thus, the generation of electron mass, the Dirac CP-violating phase and the baryon number asymmetry are closely connected in this simple scenario. Certainly, further efforts should be devoted to constructing a full model for lepton mass spectra and flavor mixing based on the $S^{}_{3{\rm L}} \times S^{}_{3{\rm R}}$ flavor symmetry. The phenomenological studies in the present paper are helpful in looking for a viable way of flavor symmetry breaking, and instructive for understanding lepton mass spectra, flavor mixing pattern and CP violation.

\section*{Acknowledgements}

This work was supported in part by the NNSFC under Grant No.11325525, by the National Recruitment Program for Young Professionals and by the CAS Center for Excellence in Particle Physics (CCEPP).

\section*{Appendix A}

In this appendix, we show explicitly how to derive the lepton mass matrices in Eq.~(4) in the limit of an exact $S^{}_{3{\rm L}} \times S^{}_{3{\rm R}}$ symmetry. First of all, let us summarize the main properties of the symmetry group $S^{}_3$ of the permutations of three objects. The order of $S^{}_3$ is equal to $3! = 6$, and all the six elements correspond to the following transformations
\begin{eqnarray}
e: &~& (x^{}_1, x^{}_2, x^{}_3) \to (x^{}_1, x^{}_2, x^{}_3) \; , \nonumber \\
a^{}_1: &~& (x^{}_1, x^{}_2, x^{}_3) \to (x^{}_2, x^{}_1, x^{}_3) \; , \nonumber \\
a^{}_2: &~& (x^{}_1, x^{}_2, x^{}_3) \to (x^{}_1, x^{}_3, x^{}_2) \; , \nonumber \\
a^{}_3: &~& (x^{}_1, x^{}_2, x^{}_3) \to (x^{}_3, x^{}_2, x^{}_1) \; , \nonumber \\
a^{}_4: &~& (x^{}_1, x^{}_2, x^{}_3) \to (x^{}_3, x^{}_1, x^{}_2) \; , \nonumber \\
a^{}_5: &~& (x^{}_1, x^{}_2, x^{}_3) \to (x^{}_2, x^{}_3, x^{}_1) \; ,
\end{eqnarray}
which can also be represented by six matrices
\begin{eqnarray}
P(e): \left(\begin{matrix} 1 & 0 & 0 \cr 0 & 1 & 0 \cr 0 & 0 & 1\end{matrix}\right) \; , \quad P(a^{}_1): \left(\begin{matrix} 0 & 1 & 0 \cr 1 & 0 & 0 \cr 0 & 0 & 1\end{matrix}\right) \; , \quad P(a^{}_2): \left(\begin{matrix} 1 & 0 & 0 \cr 0 & 0 & 1 \cr 0 & 1 & 0\end{matrix}\right) \; , \nonumber \\
P(a^{}_3): \left(\begin{matrix} 0 & 0 & 1 \cr 0 & 1 & 0 \cr 1 & 0 & 0\end{matrix}\right) \; , \quad P(a^{}_4): \left(\begin{matrix} 0 & 0 & 1 \cr 1 & 0 & 0 \cr 0 & 1 & 0\end{matrix}\right) \; , \quad P(a^{}_5): \left(\begin{matrix} 0 & 1 & 0 \cr 0 & 0 & 1 \cr 1 & 0 & 0\end{matrix}\right) \; ,
\end{eqnarray}
acting on a reducible triplet $x \equiv (x^{}_1, x^{}_2, x^{}_3)^{\rm T}$. As is well known, these six group elements can be categorized into three conjugacy classes $C^{}_1 = \{e\}$, $C^{}_2 =\{a^{}_1 a^{}_2, a^{}_2 a^{}_1\}$, and $C^{}_3 = \{a^{}_1, a^{}_2, a^{}_2 a^{}_1 a^{}_2\}$, where $a^{}_1 a^{}_2 = a^{}_4$, $a^{}_2 a^{}_1 = a^{}_5$, and $a^{}_1 a^{}_2 a^{}_1 = a^{}_2 a^{}_1 a^{}_2 = a^{}_3$ can be easily identified. The irreducible representations of $S^{}_3$ contain two singlets ${\bf 1}$ and ${\bf 1}^\prime$, and one doublet ${\bf 2}$. The representations of the singlets ${\bf 1}$ and ${\bf 1}^\prime$ are just given by their characters, while those of the doublet ${\bf 2}$ are found to be
\begin{eqnarray}
&~& D(e): \left(\begin{matrix} 1 & 0 \cr 0 & 1 \end{matrix}\right) \; , \quad \quad \quad \quad \quad \quad D(a^{}_1): \left(\begin{matrix} -1 & 0 \cr 0 & 1 \end{matrix}\right) \; , \quad \quad \quad D(a^{}_2): \left(\begin{matrix} \displaystyle \frac{1}{2} & \displaystyle \frac{\sqrt{3}}{2} \cr \displaystyle \frac{\sqrt{3}}{2} & - \displaystyle \frac{1}{2} \end{matrix}\right) \; , \nonumber \\
&~& D(a^{}_3): \left(\begin{matrix} \displaystyle \frac{1}{2} & \displaystyle -\frac{\sqrt{3}}{2} \cr \displaystyle -\frac{\sqrt{3}}{2} & - \displaystyle \frac{1}{2} \end{matrix}\right) \; , \quad D(a^{}_4): \left(\begin{matrix} \displaystyle -\frac{1}{2} & \displaystyle -\frac{\sqrt{3}}{2} \cr \displaystyle \frac{\sqrt{3}}{2} & - \displaystyle \frac{1}{2} \end{matrix}\right) \; , \quad D(a^{}_5): \left(\begin{matrix} \displaystyle -\frac{1}{2} & \displaystyle \frac{\sqrt{3}}{2} \cr -\displaystyle \frac{\sqrt{3}}{2} & - \displaystyle \frac{1}{2} \end{matrix}\right) \; .
\end{eqnarray}
For a reducible triplet $x = (x^{}_1, x^{}_2, x^{}_3)^{\rm T}$, it is always possible to find a unitary matrix $U$, which could be used to diagonalize $P(a^{}_1)$ and leads to a block-diagonal form of $P(a^{}_i)$ (for $i = 2, \cdots, 5$) in the transformed basis $x^\prime = U^\dagger x$. One can verify that the new representation matrices $P^\prime(a^{}_i) = U^\dagger P(a^{}_i) U$ turn out to be
\begin{eqnarray}
P^\prime(a^{}_i): \left(\begin{matrix} D(a^{}_i) & 0 \cr 0 & 1\end{matrix}\right) \; ,
\end{eqnarray}
for $i = 0, 1, \cdots, 5$ and $a^{}_0 \equiv e$, and that the unitary matrix $U$ is just the democratic mixing matrix $V^{}_0$ in Eq.~(5). Now that
\begin{eqnarray}
\left(\begin{matrix} x^\prime_1 \cr x^\prime_2 \cr x^\prime_3 \end{matrix}\right) = V^\dagger_0 \left(\begin{matrix} x^{}_1 \cr x^{}_2 \cr x^{}_3 \end{matrix}\right) = \frac{1}{\sqrt{6}} \left(\begin{matrix}\sqrt{3} (x^{}_1 - x^{}_2) \cr x^{}_1 + x^{}_2 - 2x^{}_3 \cr \sqrt{2} (x^{}_1 + x^{}_2 + x^{}_3) \end{matrix}\right)\; ,
\end{eqnarray}
we can identify the singlet and doublet states
\begin{eqnarray}
{\bf S} = \frac{1}{\sqrt{3}} (x^{}_1 + x^{}_2 + x^{}_3) \; , \quad {\bf M} = \left(\begin{matrix} \displaystyle \frac{1}{\sqrt{2}} (x^{}_1 - x^{}_2) \cr \displaystyle \frac{1}{\sqrt{6}} (x^{}_1 + x^{}_2 - 2 x^{}_3)\end{matrix}\right) \; .
\end{eqnarray}
The direct product of two doublets ${\bf M}^{}_1 = (b^{}_1, c^{}_1)^{\rm T}$ and ${\bf M}^{}_2 = (b^{}_2, c^{}_2)^{\rm T}$ can be decomposed into three irreducible representations
\begin{eqnarray}
\left(\begin{matrix}b^{}_1 \cr c^{}_1\end{matrix}\right)^{}_{\bf 2} \otimes \left(\begin{matrix}b^{}_2 \cr c^{}_2\end{matrix}\right)^{}_{\bf 2} = (b^{}_1 b^{}_2 + c^{}_1 c^{}_2)^{}_{\bf 1} + (b^{}_1 c^{}_2 - b^{}_2 c^{}_1)^{}_{{\bf 1}^\prime} + \left(\begin{matrix}b^{}_1 c^{}_2 + b^{}_2 c^{}_1 \cr b^{}_1 b^{}_2 - c^{}_1 c^{}_2\end{matrix}\right)^{}_{\bf 2} \; ,
\end{eqnarray}
For more details about the $S^{}_3$ symmetry group, one should be referred to recent reviews on discrete flavor symmetries and their applications in particle physics~\cite{Ishimori:2010au,Grimus:2011fk}.

Second, we explain the assignments of fermion fields and write down the Lagrangian for lepton masses under the $S^{}_{3{\rm L}} \times S^{}_{3{\rm R}}$ symmetry, following the idea of Ref.~\cite{Harari:1978yi}. All the left-handed (right-handed) fermion fields are assigned as a reducible three-dimensional representation of $S^{}_{3{\rm L}}$ ($S^{}_{3{\rm R}}$), namely, $(\ell^{}_{e{\rm L}}, \ell^{}_{\mu{\rm L}}, \ell^{}_{\tau {\rm L}})^{\rm T}$, $(e^{}_{\rm R}, \mu^{}_{\rm R}, \tau^{}_{\rm R})^{\rm T}$ and $(N^{}_{e{\rm R}}, N^{}_{\mu{\rm R}}, N^{}_{\tau {\rm R}})^{\rm T}$, and as a trivial representation of $S^{}_{3{\rm R}}$ ($S^{}_{3{\rm L}}$). The corresponding representations of $S^{}_{3{\rm L}}$ or $S^{}_{3{\rm R}}$ group elements are given in Eq.~(41). To construct an $S^{}_{3{\rm L}} \times S^{}_{3{\rm R}}$-invariant Lagrangian for lepton masses, we are only allowed to use the left-handed singlet ${\bf S}^{}_{\ell^{}_{\rm L}} \equiv (\ell^{}_{e{\rm L}} + \ell^{}_{\mu{\rm L}} + \ell^{}_{\tau{\rm L}})/\sqrt{3}$ and right-handed singlets ${\bf S}^{}_{E^{}_{\rm R}} \equiv (e^{}_{\rm R} + \mu^{}_{\rm R} + \tau^{}_{\rm R})/\sqrt{3}$ for the charged-lepton Yukawa interaction, and ${\bf S}^{}_{\ell^{}_{\rm L}}$ and ${\bf S}^{}_{N^{}_{\rm R}} \equiv (N^{}_{e{\rm R}} + N^{}_{\mu {\rm R}} + N^{}_{\tau{\rm R}})/\sqrt{3}$ for Dirac neutrino Yukawa interaction. However, for the Majorana mass term of right-handed neutrinos, the doublet ${\bf M}^{}_{N^{}_{\rm R}} = ((N^{}_{e{\rm R}} - N^{}_{\mu{\rm R}})/\sqrt{2}, (N^{}_{e{\rm R}}+N^{}_{\mu {\rm R}} - 2 N^{}_{\tau {\rm R}})/\sqrt{6})^{\rm T}$ is also available. Hence, the invariant Lagrangian with respect to an $S^{}_{3{\rm L}} \times S^{}_{3{\rm R}}$ symmetry is given by
\begin{eqnarray}
-{\cal L}^{}_\ell = y^{}_l \overline{{\bf S}^{}_{\ell^{}_{\rm L}}} H {\bf S}^{}_{E^{}_{\rm R}} + y^{}_\nu \overline{{\bf S}^{}_{\ell^{}_{\rm L}}} \tilde{H} {\bf S}^{}_{N^{}_{\rm R}} + \frac{1}{2} \left[ \alpha^{}_{\rm R} \overline{{\bf S}^{\rm C}_{N^{}_{\rm R}}} {\bf S}^{}_{N^{}_{\rm R}} + \beta^{}_{\rm R} (\overline{{\bf M}^{\rm C}_{N^{}_{\rm R}}} {\bf M}^{}_{N^{}_{\rm R}})^{}_{\bf 1}\right] + {\rm h.c.}\; .
\end{eqnarray}
Comparing between Eq.~(47) and Eq.~(2), we can obtain Eq.~(4) by identifying $c^{}_l = y^{}_l v$, $c^{}_{\rm D} = y^{}_\nu v$, $c^{}_{\rm R} = \alpha^{}_{\rm R} - \beta^{}_{\rm R}$ and $r^{}_{\rm R} = 3\beta^{}_{\rm R}/(\alpha^{}_{\rm R} - \beta^{}_{\rm R})$.

\end{document}